\documentclass{aa}  

\usepackage{caption}
\usepackage{subcaption}
\usepackage{graphicx}
\usepackage{natbib}
\bibpunct{(}{)}{;}{a}{}{,} % to follow the A&A style
\usepackage{txfonts}
\begin{document}

   \title{Boltzmann-Poisson-like\ approach to simulating the galactic halo response to satellite accretion}

   \subtitle{Dependence on the halo density profile}

   \author{G. Aguilar-Arg\"uello\inst{1}, O. Valenzuela\inst{1},
          \and
          A. Trelles\inst{1}
          }

   \institute{Universidad Nacional Autónoma de México, Instituto de Astronomía, AP 70-264, CDMX  04510, México\\
              \email{gaguilar@astro.unam.mx, octavio@astro.unam.mx}
             }

%  \date{Received July 15, 2021; accepted July , 2021}

  \abstract
   {Recent studies have reported the detection of the galactic stellar halo wake and dipole triggered by the Large Magellanic Cloud (LMC), mirroring the corresponding response from dark matter (DM).  These studies open up the possibility of adding constraints on the global mass distribution of the Milky Way (MW), and even on the nature of DM itself, with current and upcoming stellar surveys reigniting the discussion on response modes in dynamical friction. However, the simulation of such features remains computationally challenging.}
  {Using a continuous medium approach, we investigate the density and velocity response modes in simulations of Galactic-type DM halos accreting LMC-sized satellites, including the dependence on the halo density profile.}
  {We used, for the first time in the context of galactic dynamics, a collisionless Boltzmann equation (CBE)+Poisson solver based on an existing method from the literature. We studied the dynamical density and velocity response of halos to sinking perturbers.}
  {We successfully captured both the local wake and the global over- and underdensity induced in the host halo. We also captured the velocity response. In line with previous studies, we find that the code can reproduce the core formation in the cuspy profile and the satellite core stalling. The angular power spectrum (APS) response is shown to be sensitive to each density profile. The cored Plummer density profile seems the most responsive, displaying a richness of modes. At the end of the simulation, the central halo acquires cylindrical rotation. When present, a stellar component is expected to behave in a similar fashion.}
  {The CBE description makes it tenable to capture the response modes with a better handling of noise in comparison to traditional \emph{N}-body simulations. Hence, given a certain noise level, BPM has a lower computational cost than \emph{N}-body simulations, making it feasible to explore large parameter sets. We anticipate that stellar spheroids in the MW or external galaxies could show central cylindrical rotation if they underwent a massive accretion event. The code can be adjusted to include a variety of DM physics.}

\titlerunning{Boltzmann Simulation of Halo Response}
\authorrunning{G. Aguilar-Arg\"uello}

   \maketitle
%
%-------------------------------------------------------------------

\section{Introduction} \label{sec:intro}
The current paradigm assumed for cosmic structure formation is the $\Lambda$ cold dark matter ($\Lambda$CDM) model. This scenario encompasses collisionless cold Dark Matter (DM) and a cosmological constant \citep{2020A&A...641A...6P}. The dark halo structure of galaxies has been extensively studied in $\Lambda$CDM-related models, most frequently with rotation curves, disk dynamics, or galaxy scaling relations \citep{2016MNRAS.462..893R,2018MNRAS.479.2133A}. The dynamics of satellite galaxies are also critical tests \citep{2018MNRAS.476.5669W}, particularly for nearby galaxies \citep{2020MNRAS.494.4291C}. Recent and upcoming surveys such as GAIA \citep{2021A&A...649A...1G}, APOGEE \citep{2017AJ....154...94M, 2020A&A...644A..83F}, DESI \citep{2021AAS...23730306C}, WEAVE \citep{2012SPIE.8446E..0PD}, and LSST \citep{2018IAUS..334..233R} will produce an exquisite mapping of the Milky Way's (MW) stellar halo. In addition, these surveys will open up an opportunity to track the subtle stellar response that is presumably paired with the DM halo response, constraining such processes as dynamical friction as well as the properties of halos, or even those of DM itself, as recently shown via observations by \citet{2021Natur.592..534C}. Using high-resolution \emph{N}-body simulations, \citet{2021ApJ...919..109G} investigated the halo response modes to satellite accretion following such pioneering studies as \citet{1989MNRAS.239..549W}. In addition, \cite{2016MNRAS.457.2164O,2021ApJ...916...55T} used super high-resolution simulations to study the relative importance of global mode and local wake in dynamical friction.

The recent studies mentioned above \citep{2016MNRAS.457.2164O,2021ApJ...919..109G,2021ApJ...916...55T} have revealed a rich response mode population in fully self-consistent calculations reaching particle numbers of around $10^8 - 10^9$  based on large computational resources. However, exploring a large parameter space using \emph{N}-body simulations, while possible, may prove challenging due to the large amount of computational resources required. Therefore, devising an alternative technique may be useful. In the field of numerical simulations, testing results using different techniques consistently contributes to their individual robustness. Additionally, a continuum medium approach to dynamical problems is always interesting thanks to shot-noise handling, which is critical when the main mechanism we aim to capture is low-amplitude overdensities, as touched upon in the  discussion above.

In this work, we propose using a flexible alternative or complementary strategy based on a continuous medium description for matter density. Our code is based on recent Collisionless Boltzmann Equation (CBE) solver implementations aimed at studying cosmic neutrinos \citep[][hereafter BD16]{2016JCAP...11..015B}. In cosmological simulations with massive neutrinos, the shot-noise density fluctuations are comparable to the model's actual initial inhomogeneities, triggering artificial power. Traditionally, this is alleviated thanks to the use of a large number of particles, but a continuous medium description such as BD16 is an interesting choice. The BD16 method solves the Boltzmann moment equations on a grid and uses tracer particles to estimate the higher-order velocity moments of the Boltzmann hierarchy (instead of the density). We illustrate the code's performance by tracking isolated halo and satellite sinking simulations.

In Section \ref{sec:BPM}, we describe the general aspects of the BPM code. Section \ref{sec:Sims} presents the simulations following the wake, the global density, and the kinematics response in sinking satellite experiments exploring the dependence of different host halo density profiles. Our results are presented in Section \ref{sec:Results}, including comparisons to \emph{N}-body codes (Section \ref{sec_BPMvsNbody}). Finally, our discussion and conclusions are given in Section \ref{sec:Discusion}.

\section{BPM code} \label{sec:BPM}
Solving the  CBE+Poisson equations directly allows us to address many important aspects of the dynamics of self-gravitating systems by maintaining the advantage of a continuous media description while avoiding particle shot-noise. However, such an approach is quite expensive due to the high dimensionality of the phase-space, however the  world's top supercomputers could make it feasible, as, for instance, illustrated in recent works by \citet{2013ApJ...762..116Y,2020ApJ...904..159Y}. To reduce the cost compared to the direct approach, we propose, instead (following BD16) to use a moment hierarchy approach. Such a technique allows for a 3D description that is more manageable from the computational point of view because it reduces aspects of dimensionality, as compared to a full Vlasov one. We calculate the first two CBE moments or the continuity and Euler-like equations coupled with the Poisson equation on a fixed 3D Cartesian grid. The order of the moment expansion may be questioned, in particular, crossed terms are not included. However, in contrast to self-consistent field codes \citep[e.g.,][]{1989MNRAS.239..549W}, our approximation is not directly performed on density or gravitational potential. As a healthy measure, we present in Section \ref{sec_BPMvsNbody} a quantitative comparison to full \emph{N}-body codes such as a fast multipole method \citep[FMM,][]{1987JCoPh..73..325G,2000ApJ...536L..39D} and a particle mesh \citep[PM,][]{1988csup.book.....H} ones.

The set of equations that we solve are listed below:
\begin{align}
    \frac{\partial \rho}{\partial t} + \frac{\partial (\rho v^i)}{\partial {x^i}}  = 0 \label{Cont} ,\\ 
    \frac{\partial (\rho v^i)}{\partial {t}} + \frac{\partial (\rho v^i v^j)}{\partial {x^j}} = -\rho \frac{\partial \Psi }{\partial x^i} - \frac{\partial (\rho \sigma^{ij})}{\partial {x^j}} \label{Euler},\\ 
    \nabla^2\Psi  = 4 \pi G \rho. \label{poiss}
\end{align}

This set is equivalent to what is included in the cosmological case presented by BD16. A critical point when using a moment expansion approach is the closure condition used to truncate the expansion. Following BD16, we close the CBE hierarchy by using test particles to estimate moments of the distribution function, such as the velocity dispersion tensor and the bulk velocity. We use, at all times, the continuous density from the CBE moments solution to solve the Poisson equation (Eq. \ref{poiss}). Additionally, we evolved in time the test particles using the leap-frog integrator (Kick-Drift-Kick, KDK) and the gravitational potential calculated from the Poisson equation. Then, we use the trajectories of the test particles to calculate the velocity dispersion tensor and the bulk velocity to close the CBE hierarchy. It is important to say that such kinematics estimation is the principal noise source; all the other quantities are computed at the continuous medium regime.

To solve the coupled moment equations (Eqs. \ref{Cont} and \ref{Euler}), following the BD16 implementation, we first use the directional splitting technique \citep{2012nmhe.book...87T}. We then use a 1D piecewise linear advection solver, which is non-linear in space. In addition, we use the Superbee flux delimiter \citep{1983JCoPh..49..357H}, which has minimum diffusivity. The CBE moments are evolved in time by a total variation diminishing (TVD) Runge-Kutta on the order of two (RK2) integrator, which is known to preserve the TVD properties \citep{2003PASP..115..303T}. Therefore, our code (to which we  refer to as BPM) is stable to artificial oscillations since it is TVD. Based on the previous discussion, the number of operations in BPM is similar to that of the PM code, exchanging the CIC cost for the advection part, which makes BPM more expensive than a PM. 

Our implementation adds the possibility of following two separated species, either both of the species followed by the CBE moments (dubbed as CBE+CBE) or one species followed by CBE and the other by PM\footnote{Instead of the PM code, another \emph{N}-body code may be used} (hereafter CBE+PM). In the CBE+PM case, the species are coupled only through the Poisson equation. This capability may be useful in attempting to follow a stellar component in addition to the DM density; however, in this paper, we are only considering the DM component. The CBE+PM version is also useful because it allows for savings in computational cost, as compared to the CBE+CBE version.

\begin{table}
    \caption{Mass and scale radius of the halo host models.}
    \label{table:1}
    \centering
    \begin{tabular}{ c c c }
        \hline\hline
        \textbf{Host} & \textbf{Mass ($10^{12} M\odot$)} & \textbf{$R_s$} (kpc)\\
        \hline
        Plummer & 1.00 & 30\\
        NFW1 & 0.98 & 52\\
        NFW2 & 0.80 & 30\\
        \hline
    \end{tabular}
    \tablefoot{The quoted mass is the one inside 100 kpc. To generate the NFW models, we included a \emph{sech} (hyperbolic secant) weighting function that produced a smoother cut-off at large radii, $r_\mathrm{trunc}$, as in \citet{2005MNRAS.363.1057D} models.}
\end{table}

\begin{figure}
    \centering
    \includegraphics[trim=0.3cm 0.1cm 1.2cm 1.2cm, clip, width=0.43\textwidth]{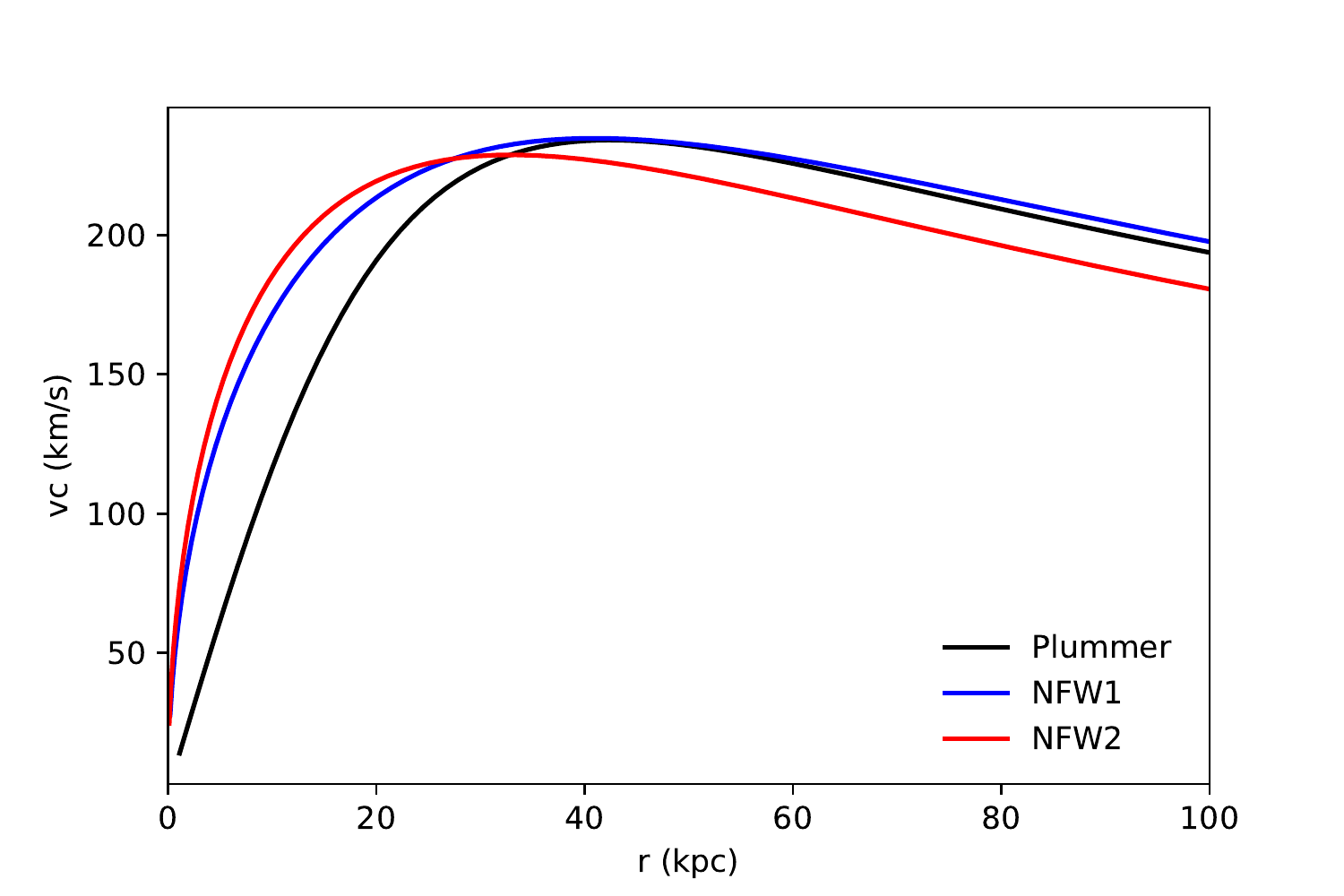}
    \caption{Dark halos' circular velocity curve. NFW1 (blue) has a similar mass as the Plummer (black) model and also the same circular velocity out of 30 kpc and NFW2 (red) is $20\%$ less massive but more concentrated and with a similar $V_\mathrm{max}$.\label{fig:Vc} }
\end{figure}

\begin{figure}
    \centering
    \includegraphics[width=0.4\textwidth]{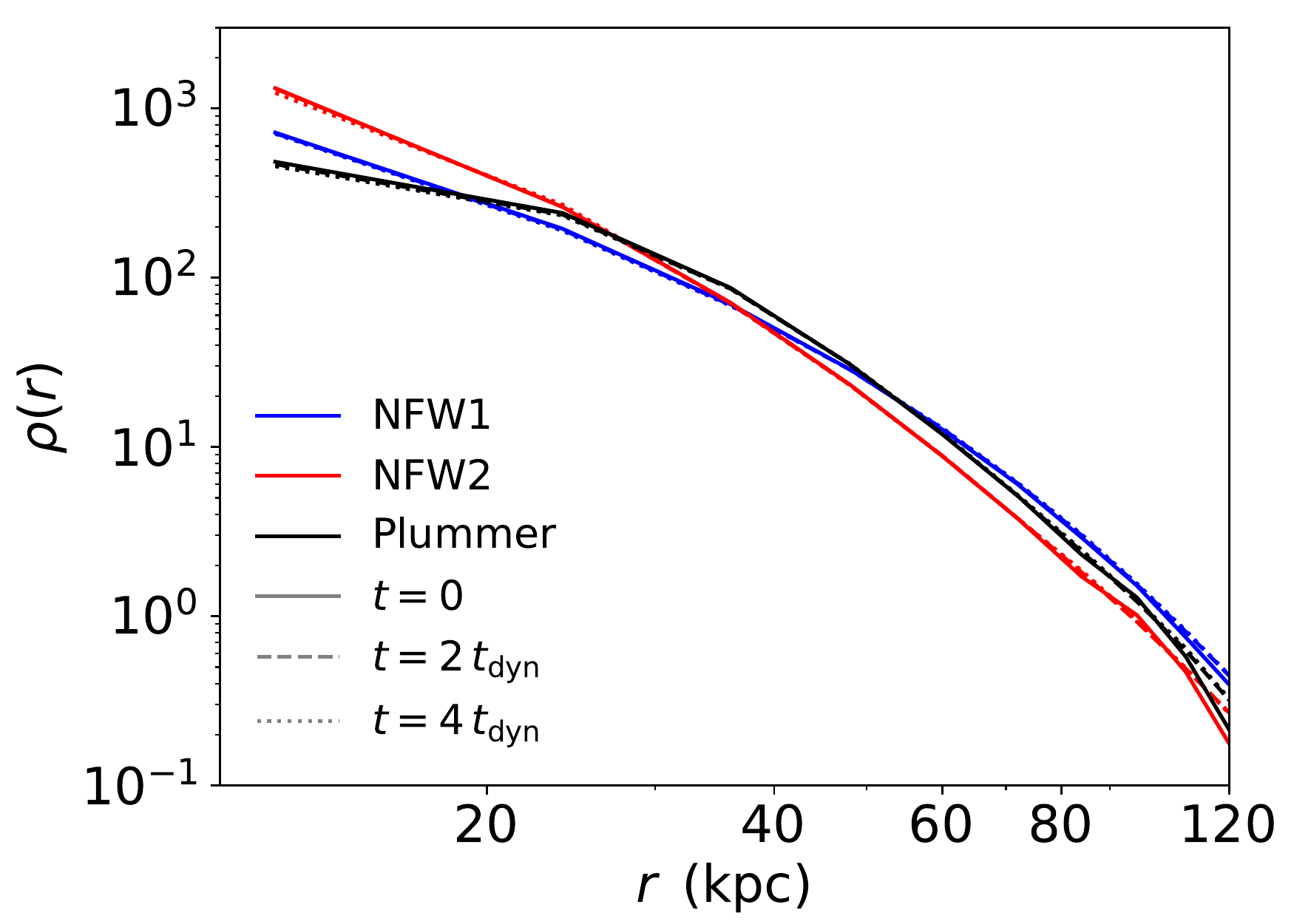}
    \caption{Stability test for BPM. We present the density profile for each of the three halo models at different times up to $4.0\,t_\mathrm{dyn}$. The density profile inside 100 kpc barely changes, thus confirming that the BPM accurately calculates its evolution. \label{fig:evol_dens_HA} }
\end{figure}

\begin{figure*}
    \centering
    \includegraphics[width=0.8\textwidth]{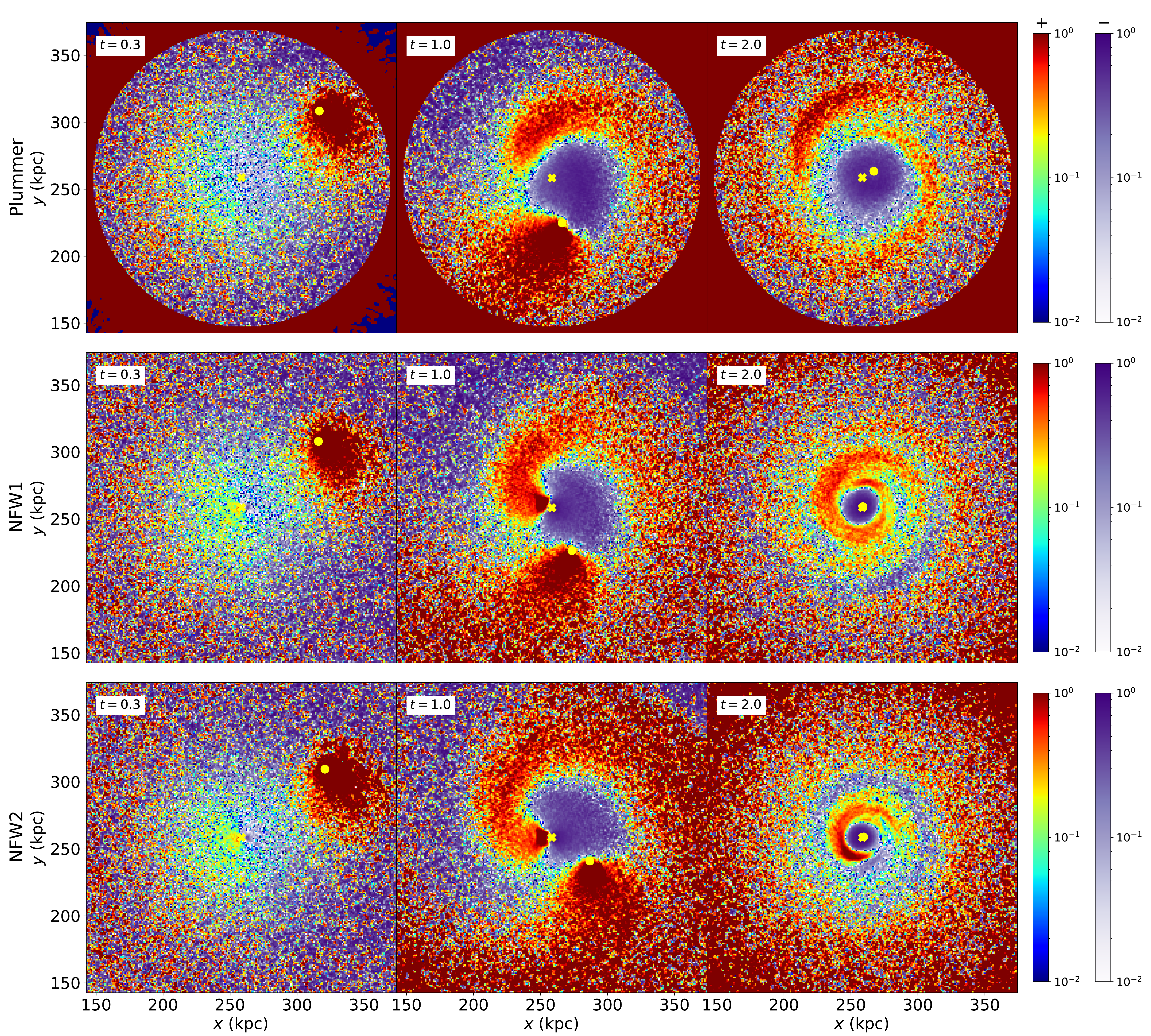}
    \caption{Evolution of halo overdensity ($\rho_t/\rho_0 - 1$) maps for the three host halos shown in Figure \ref{fig:Vc}. Purple colors are for negative overdensity values and red for positive. The upper row corresponds to the Plummer halo model, the middle row to NFW1 (same $V_c$ as the Plummer model out of 30 kpc), and the bottom row to the smaller NFW2 mass model with higher concentration. Each column corresponds to 0.3 (left), 1.0 (center), and 2.0 (right) dynamical times ($t_\mathrm{dyn}$), respectively. The host center and the satellite positions are marked by a yellow cross and solid circle, respectively. The wake and the global response are evident in all panels. The Plummer cored model starts to deviate significantly once the core radius is reached by the satellite. There are small quantitative differences in the wake and dipole for each model. At 2.0 $t_\mathrm{dyn}$, the global anisotropy is smaller, but other modes are visible. \label{fig_odensity} }
\end{figure*}

\begin{figure}
    \centering
    \includegraphics[width=0.4\textwidth]{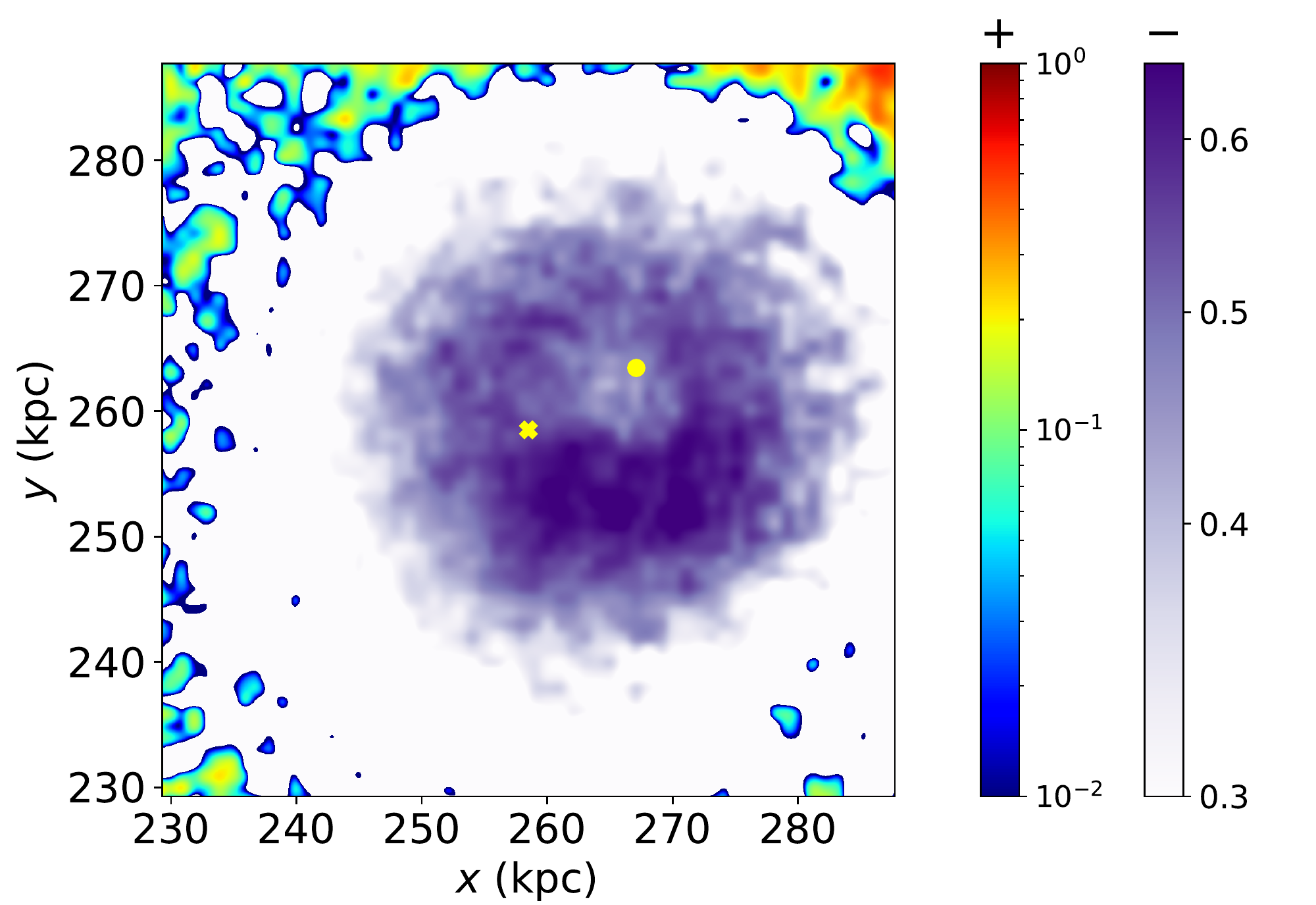}
    \caption{Modes in over- and underdensity zoom map inside the core radius for the Plummer model. We can observe a small overdensity leading the satellite (cross) and also a trailing underdensity. \label{fig_Zoom} }
\end{figure}

\begin{figure}
    \centering
    \includegraphics[width=0.45\textwidth]{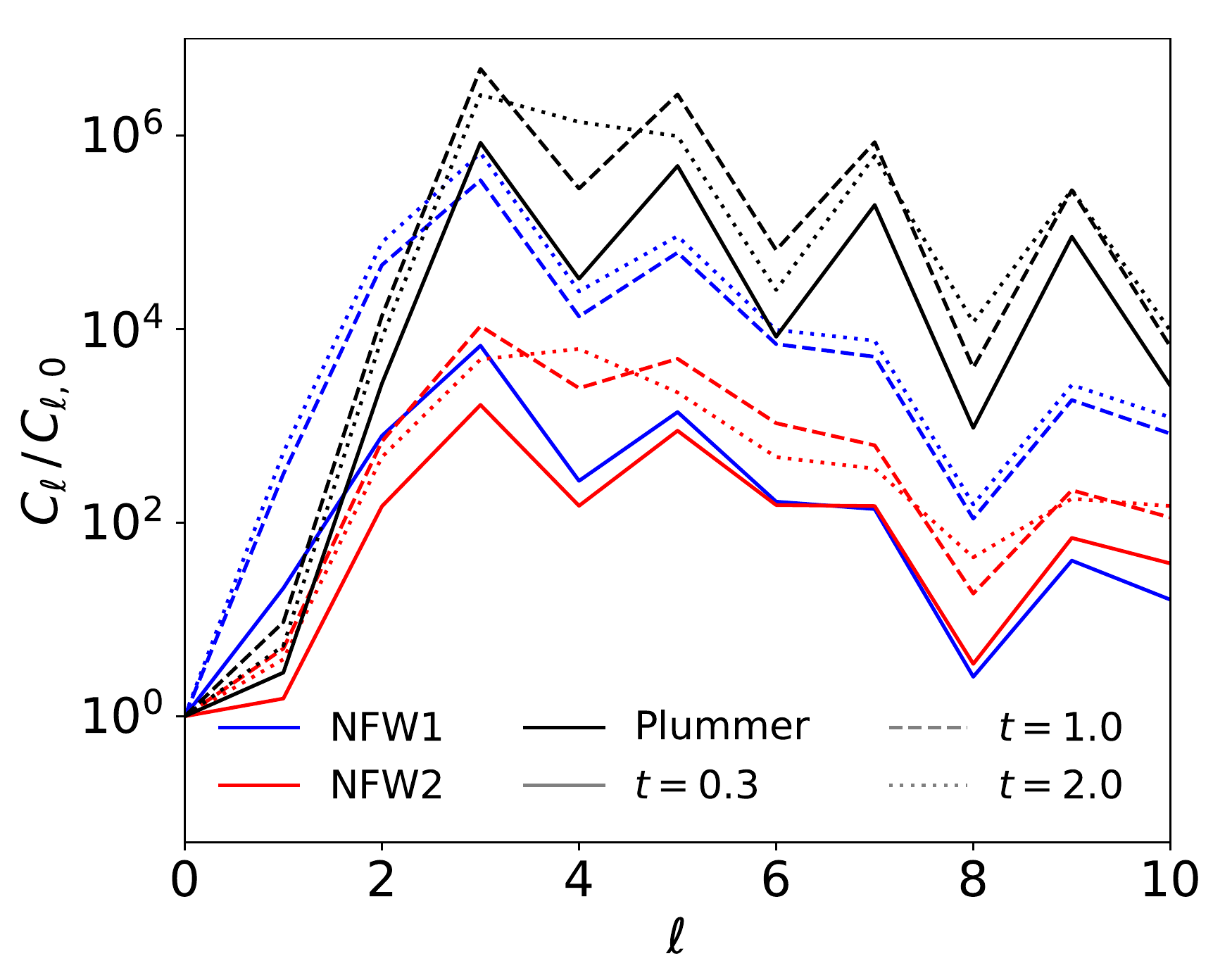}
    \caption{Evolution of halo density angular modes for the three halo models calculated with the BPM code. The peak at $\ell=3$ (see footnote \ref{APS_DensPart} for consistency with previous works) has the largest amplitude for Plummer (black) and NFW1 (blue) models at all times. However, the NFW2 (red) model shows an increasingly flatter spectrum. This is consistent with maps shown in Figure \ref{fig_odensity}, where the dipole has a comparable amplitude with the wake. \label{fig_APS_Evol} }
\end{figure}

\section{Simulations}  \label{sec:Sims}
To study the response of the underlying DM halo to a satellite accretion, we used spherical halos without any loss of generality. Since the typical particle speed is small compared with the speed of light, we only solve the continuity-like equation (1st CBE moment). This has been illustrated in BD16  in reference to their isolated Plummer halo case and cosmological CDM experiments. For the purposes of analysis, it is convenient to simulate the host halo and satellite as different species. As mentioned previously, BPM has two versions for simulations with two species: CBE+CBE and CBE+PM. We performed tests using both versions of the code; in particular, for the CBE+PM version, we evolved the host halo using CBE and the satellite using PM. We found that both CBE+CBE and CBE+PM gave similar results regarding the satellite sinking history; however, the CBE+PM is less computationally expensive. For this reason, the results that we present are based on the CBE+PM version of the code.

\subsection{Models and initial conditions} \label{subsec:models}
The host halo configuration is either a Plummer \citep{1911MNRAS..71..460P} density profile or a cuspy NFW \citep{1997ApJ...490..493N} model, with spherical symmetry in both cases. For the satellite, we consider only a Plummer density profile. The satellite and the halo have the same mass resolution. The initial conditions were generated by sampling the Plummer distribution function and the NFW model using the publicly available MakeHalo code \citep{2007MNRAS.378..541M}. The mass ratio between the satellite and host is 10$\%$. We placed the satellite at a distance of 90 kpc with an orbital circularity of 0.8. Table \ref{table:1} presents the parameters of the models. We generated the initial conditions for the BPM code directly using the analytical density solution from the equations in \citet{2007MNRAS.378..541M}. It is important to state that we did not pretend to simulate a specific galaxy model. We used three distinct halos with $V_\mathrm{max}$ close to 220 km/s but with different density profiles; a Plummer model, an NFW with a similar circular velocity out of 30 kpc (NFW1), and a 20$\%$ less massive NFW halo with a higher concentration but with the similar maximum circular velocity (NFW2), as seen in Figure \ref{fig:Vc}. The satellite is represented by a Plummer model with a core radius of 3.6 kpc. Before starting the merger simulation, we evolved all the halos in isolation for 2.0 dynamical times, $t_\mathrm{dyn}$.

\begin{figure*}
    \centering
    \includegraphics[width=0.66\textwidth]{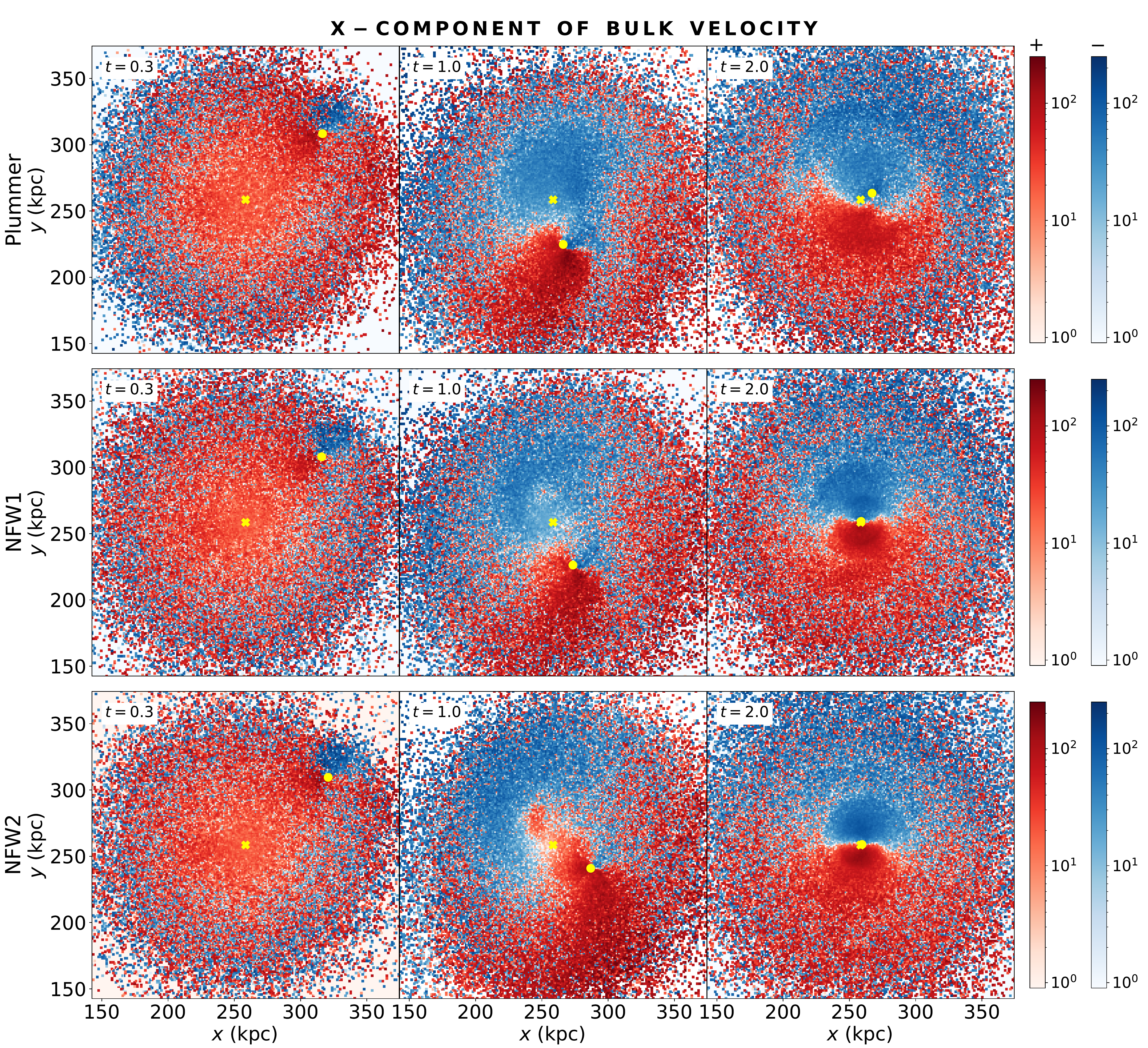}\hfill
    \includegraphics[width=0.66\textwidth]{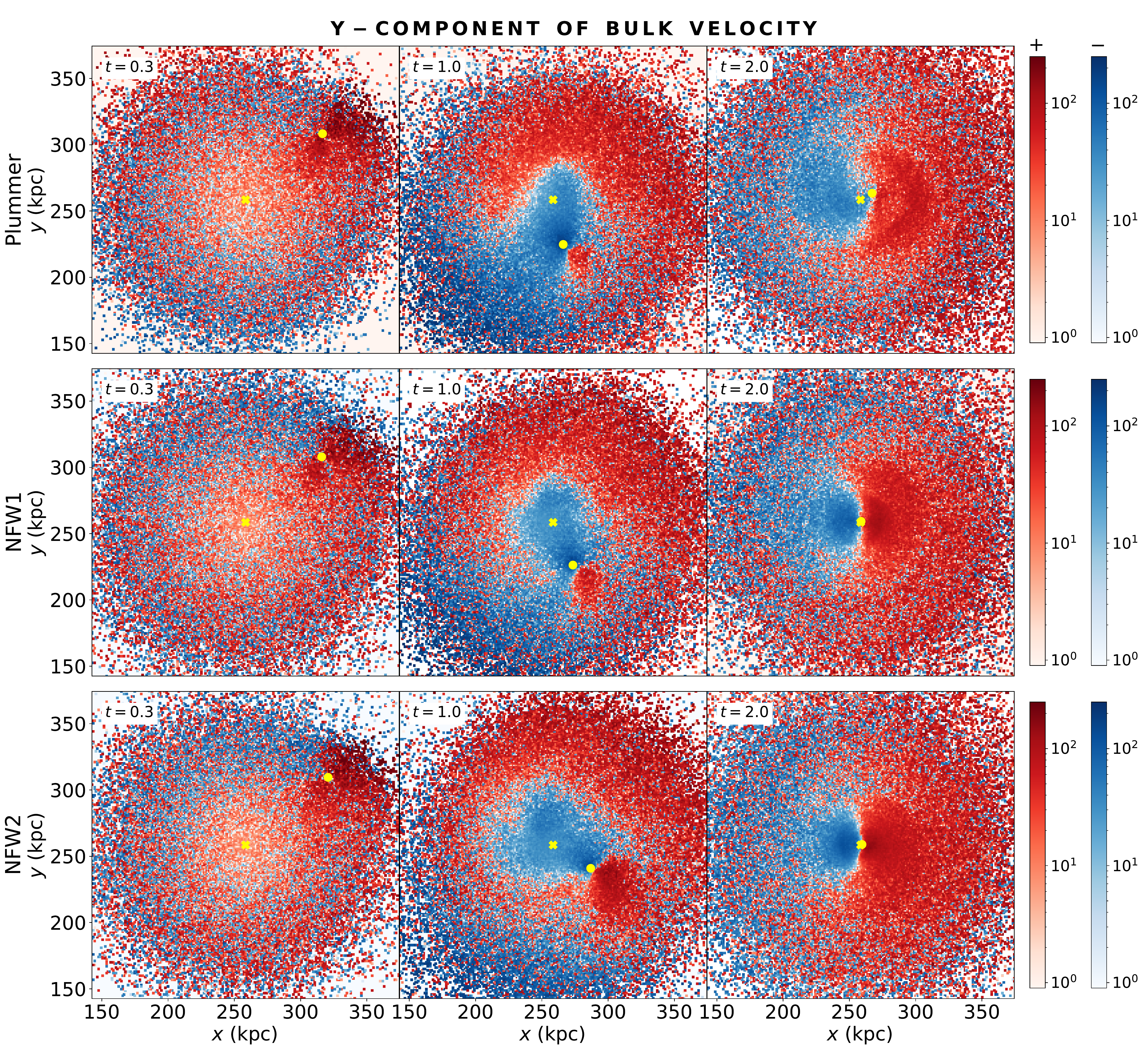}\hfill
    \caption{Velocity maps. The map shows the $X$ (upper) and $Y$ (lower) bulk velocity components (positive red, negative blue). As in Figure \ref{fig_odensity}, the columns correspond to, from left to right, $t=0.3,1.0$, and $2.0\,t_\mathrm{dyn}$, respectively. The rows from top to bottom refer to the Plummer, NFW1, and NFW2 halos, respectively. Satellite and host centers are pointed by a yellow circle and a cross, respectively. The halo response behind the satellite is clearly seen in the first two columns. The map for NFW2 starts to deviate from the other two models at $t=1.0$, suggesting that the momentum redistribution is different. The differences between the middle and right panels, for all the models, suggest that angular momentum transfers from the satellite to the background halo. \label{fig_Velocity} }
\end{figure*}

\begin{figure}
    \centering
    \includegraphics[width=0.49\textwidth]{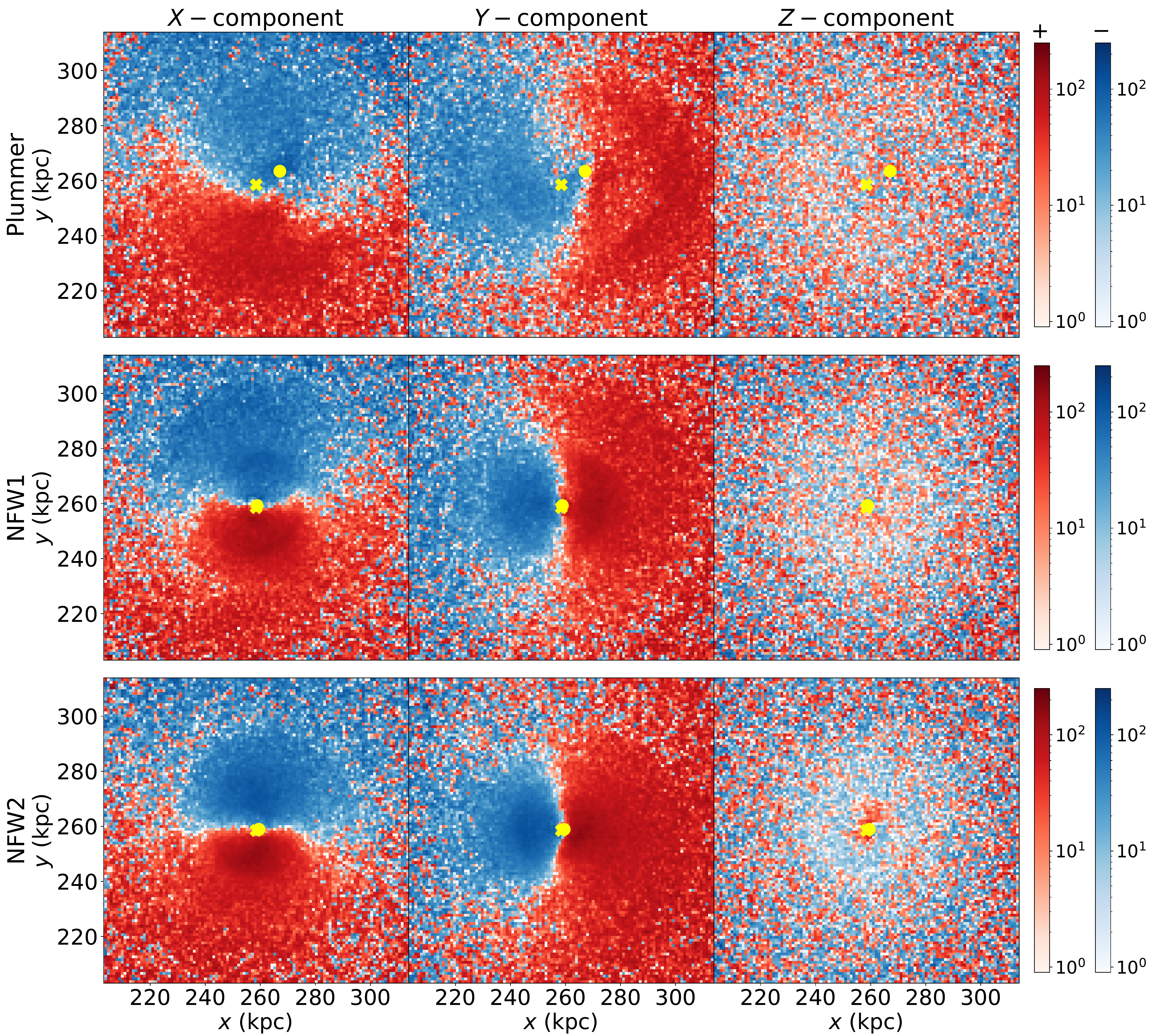}\hfill
    \caption{Central kinematics at the end of the simulation for the three simulations models. Each row corresponds to a halo model, from top to bottom: Plummer, NFW1, and NFW2. The columns from left to right show the $X$, $Y$, and $Z$ bulk velocity components, respectively. The first evident feature is that all the angular momentum exchange happens in the orbital plane ($X-Y$); therefore, the $Z$ velocity component shows no pattern. Nearly cylindrical rotation is evident in the simulations' final stage. \label{fig_CStalling.pdf} }
\end{figure}

\begin{figure}
    \centering
    \includegraphics[width=0.42\textwidth,angle=0]{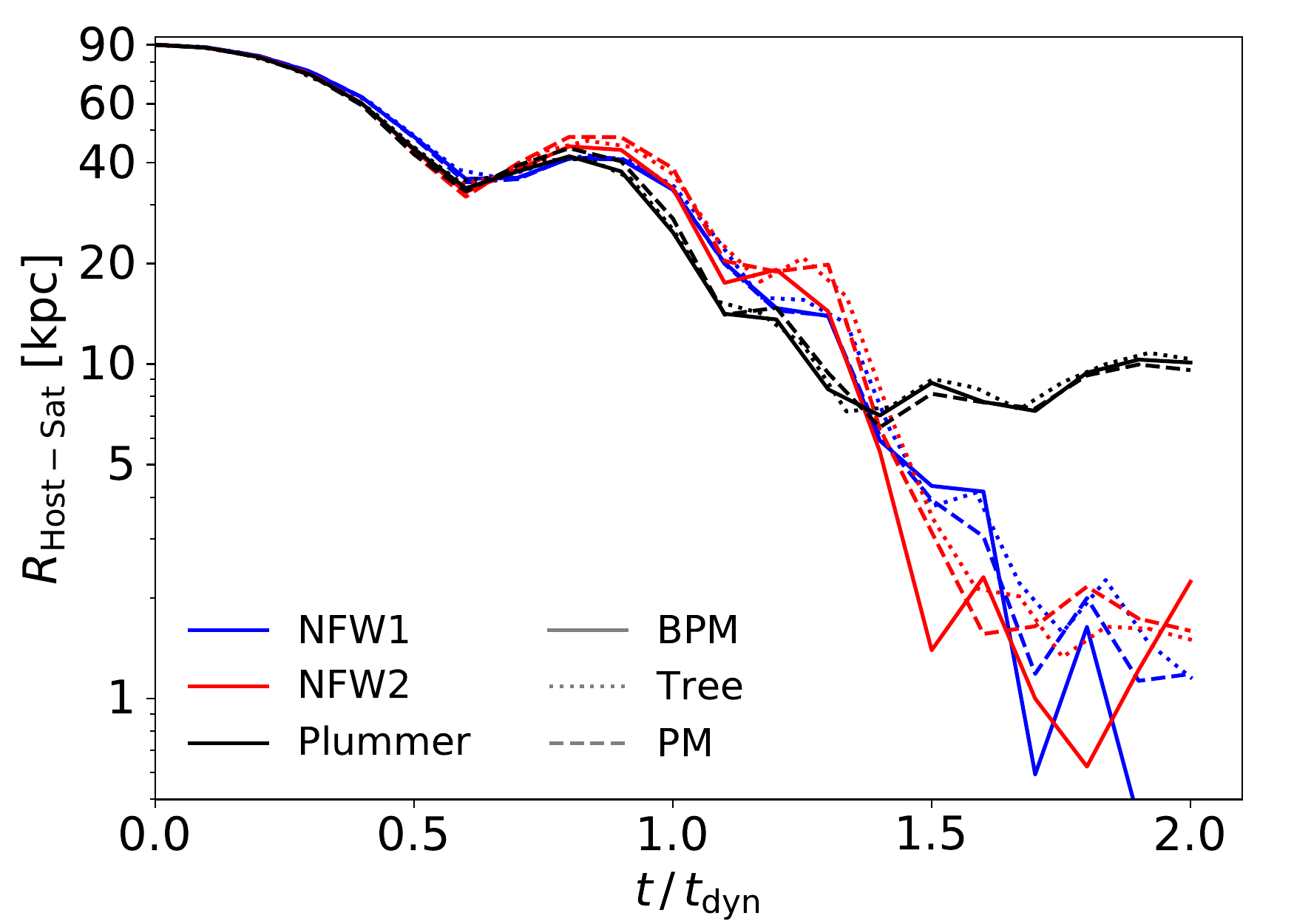}
    \caption{Sinking satellite history for different models and codes. Solid, dotted, and dashed lines correspond to BPM, Tree, and PM codes, respectively. The three codes are consistent with each other up to $t=1.3\,t_\mathrm{dyn}$ when small differences start to appear. The blue, red, and black colors correspond to NFW1, NFW2, and Plummer models simulations, respectively. The core stalling of the sinking satellite is evident in the Plummer model simulation. \label{BPMTEST} }
\end{figure}

\begin{figure}
    \centering
    \includegraphics[width=0.4\textwidth]{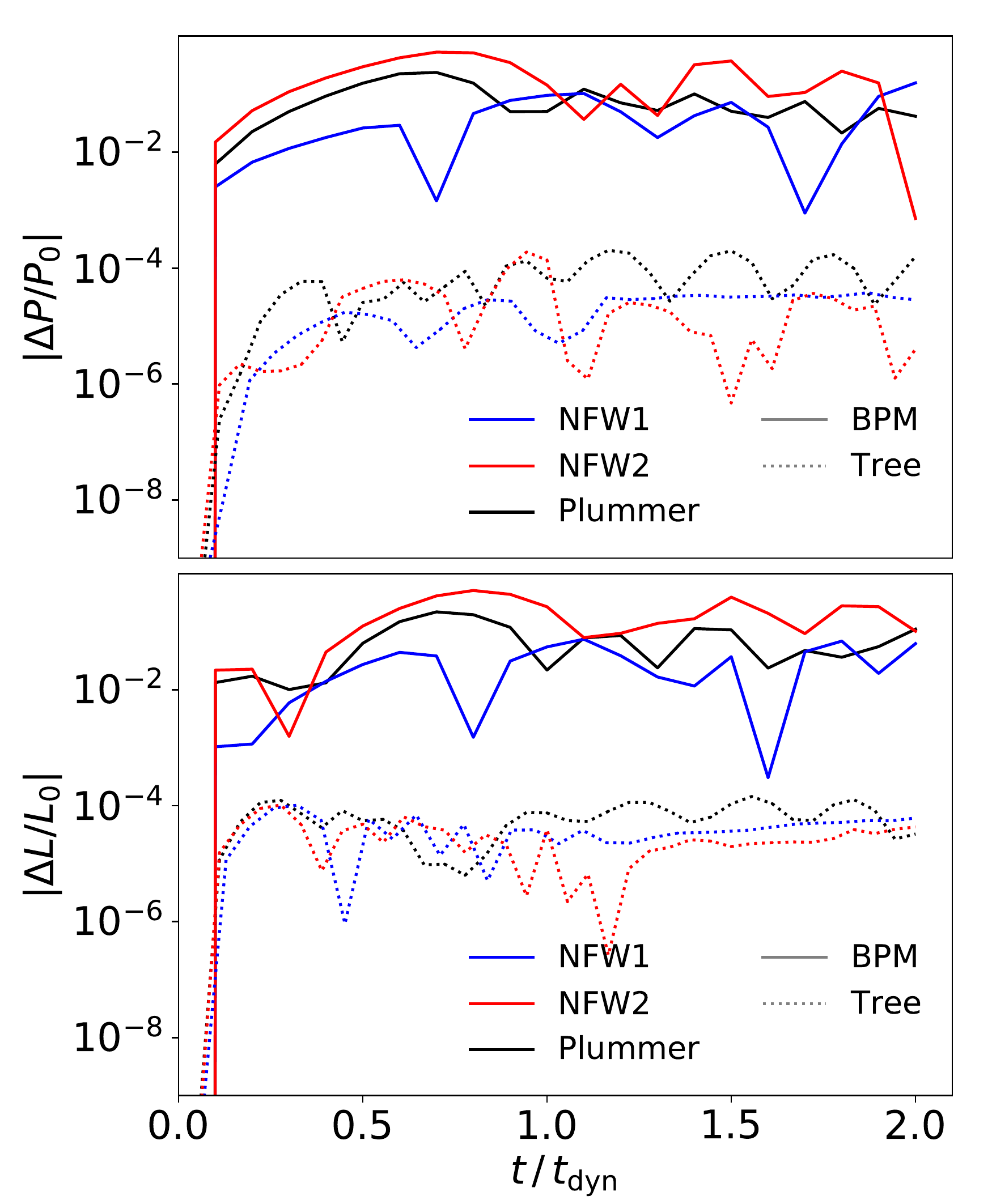}
    \caption{Momentum conservation. The upper panel shows the linear momentum evolution for the three halos with two different codes: BPM (solid lines) and full \emph{N}-body Tree code (dotted lines). The lower panel shows the angular momentum conservation for the three models with the corresponding codes. \label{Code_Momentum} }
\end{figure}

\begin{figure}
    \centering
    \includegraphics[width=0.49\textwidth]{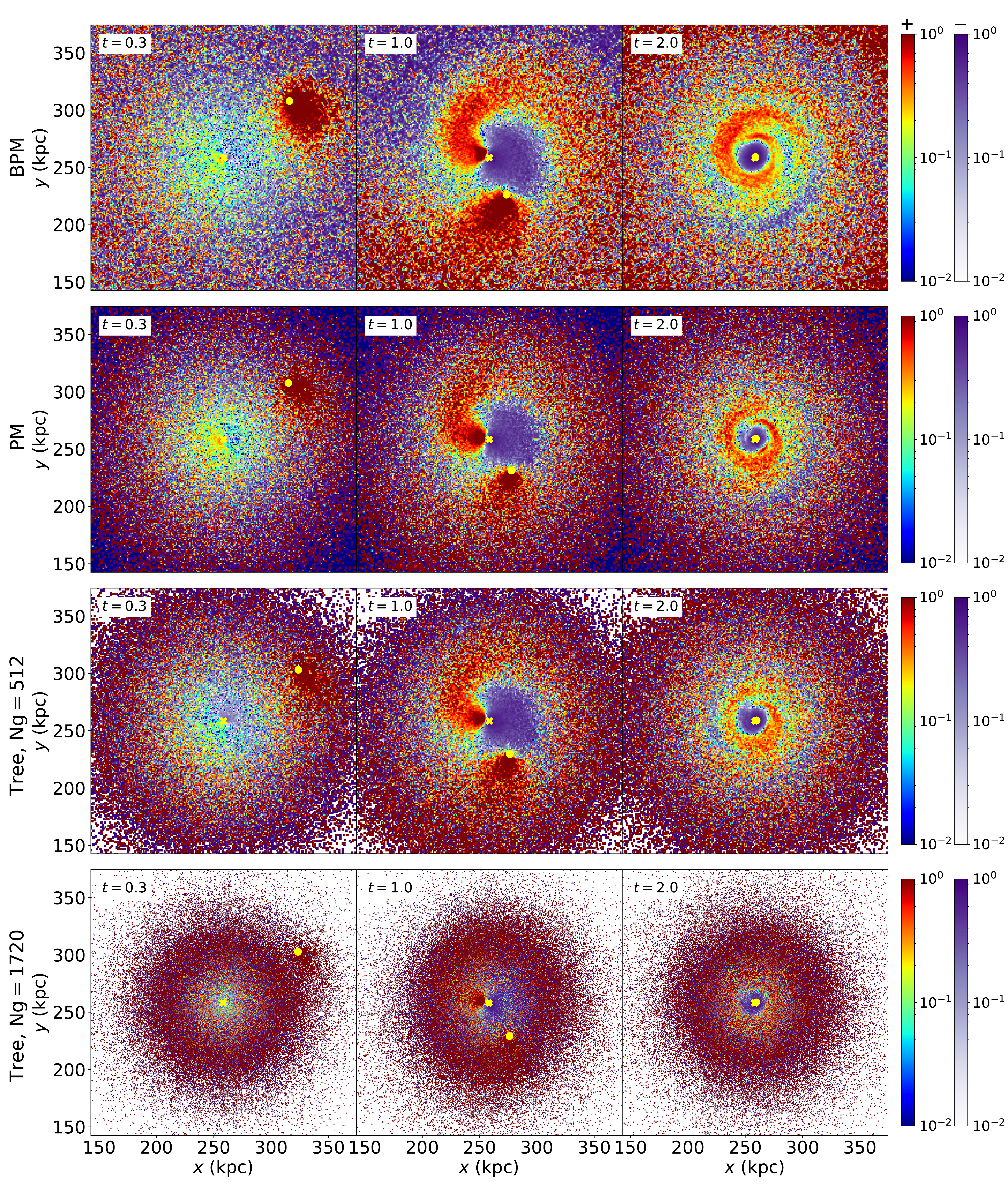}
    \caption{Overdensity maps noise effects with different codes and smoothing. The top row corresponds to BPM, where the overdensity is calculated directly with the solution of Continuity+Poisson equations. The second row shows the PM overdensity after applying CIC to the particle distribution. The two lowest rows correspond to the Tree code after applying CIC, using a grid resolution of 1.0 ($512^3$ cells) and 0.3 ($1720^3$ cells) kpc, respectively. The highest resolution hardly shows the wake and global response. However, the smoothed case shows a host halo response close to the PM case, demonstrating a dependence on the grid smoothing. \label{Code_SN} }
\end{figure}

\begin{figure}
    \centering
    \includegraphics[width=0.44\textwidth]{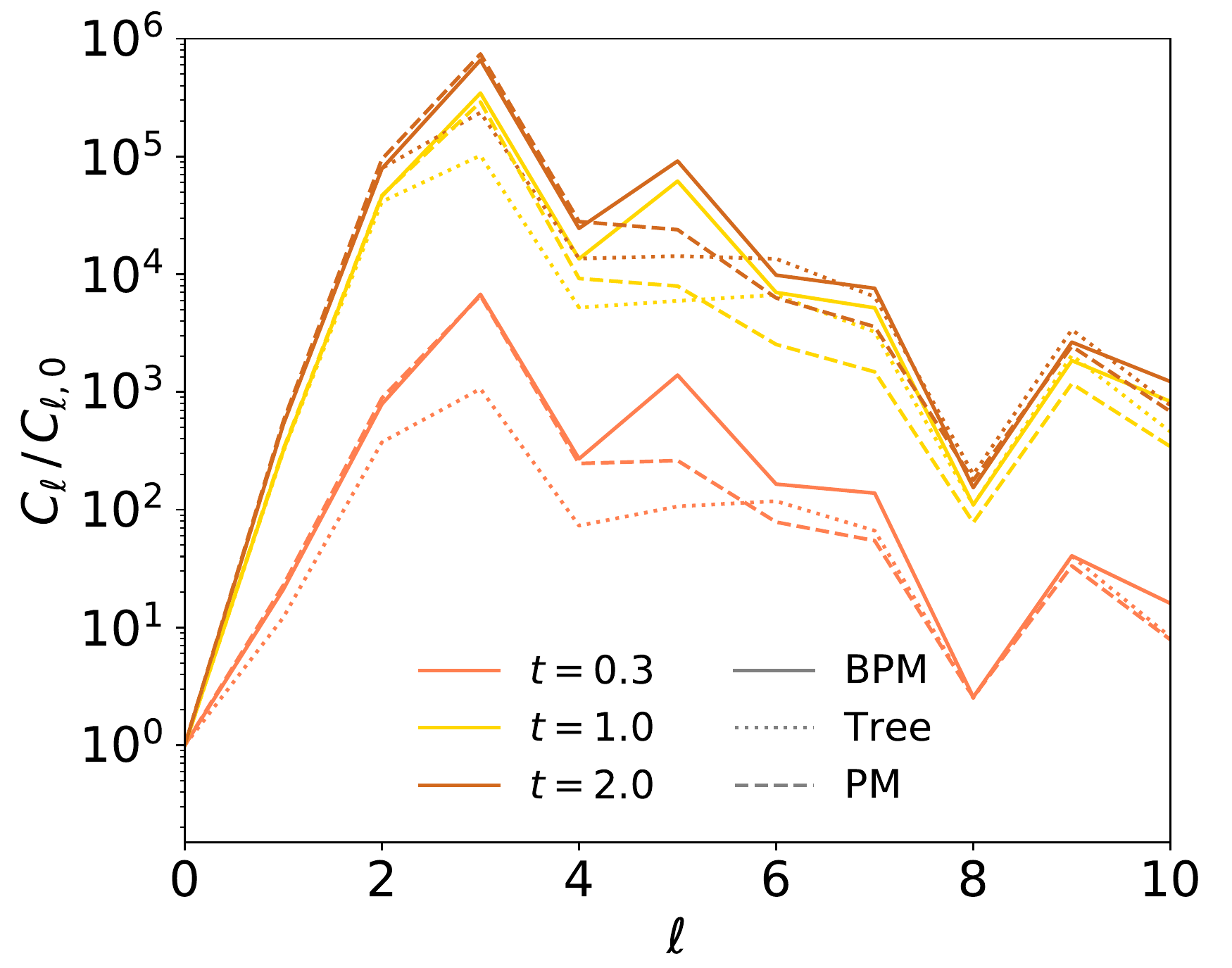}
    \caption{Evolution of density angular modes for the NFW1 halo with different codes. The peak at $\ell=3$ has the largest amplitude at any time. It has the same amplitude for BPM (solid line) and PM (dashed line), whereas it has a lower amplitude for the Tree code (dotted line). The amplitude of the peaks at $\ell=5$ and $\ell=7$ differs between codes. \label{fig_APS} }
\end{figure}

\begin{figure}
    \centering
    \includegraphics[width=0.4\textwidth]{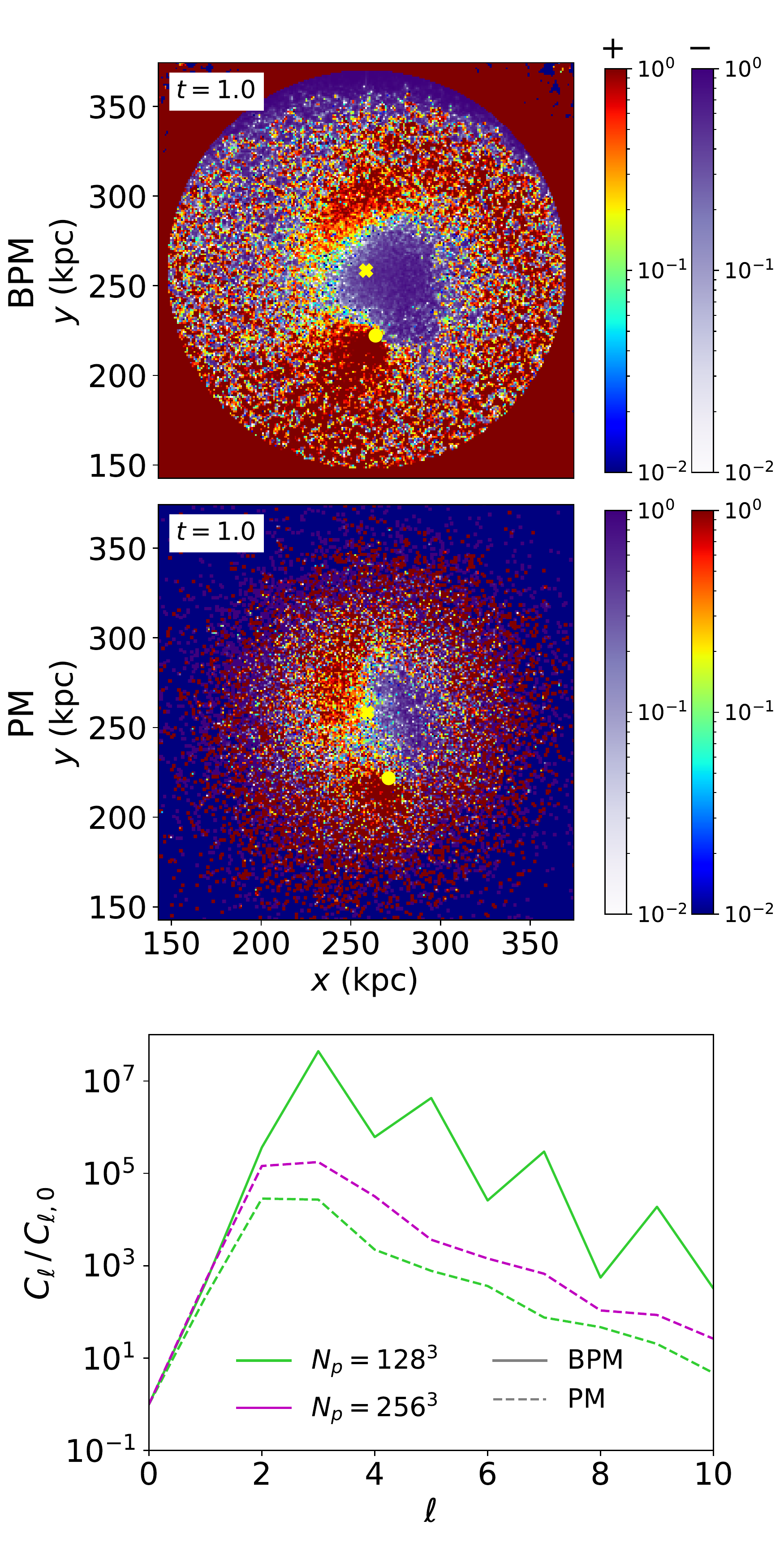}
    \caption{Noise damped modes for the Plummer model. The upper panels show the overdensity maps at the orbital plane using BPM (top) and PM (middle). The wake and the overall map are affected by noise in the PM test. The bottom panel shows the APS (green lines), where the $\ell=3$ peak is severely damped for the PM code (dashed line). Both codes use $512^3$ cells, $128^3$ particles, and a spatial resolution of 1 kpc. Additionally, we include a PM case using $256^3$ particles (magenta dashed line); the APS amplitude is greater than the previous test. The dependence on particle number suggests that there is a shot-noise effect in the PM APS. \label{fig_NoiseComp} }
\end{figure}

\subsection{Numerical experiments}
We performed all calculations on a grid of $512^3$ cells and, unless explicitly stated, $256^3$ test particles. The time step was selected according to particles velocity dispersion or sound speed for continuous medium and stability analysis (BD16). We adopted 1 kpc as the spatial resolution for the BPM and PM codes, and 0.3 kpc for the Tree code. The first set of tests followed the stability of the isolated halo for each of the halo models with the BPM code. Figure 2 shows that BPM held the density profile of all the three halos for $4.0\,t_\mathrm{dyn}$. Next, we performed the sinking satellite experiments. As expected, the satellite sinks due to dynamical friction. As expected, the cored Plummer profile shows the satellite core stalling \citep[see][and references therein]{2015MNRAS.454.3778P}, and the NFW1 and NFW2 models develop a core formation at the end of the run. More details of this process are presented in the following sections.

\section{Results} \label{sec:Results}
\subsection{ Discussion of the results} 
It is critical to accurately find the host and satellite centers since a defect in such an estimation would ultimately trigger a false density response. We estimate such positions using the shrinking-sphere method \citep{2003MNRAS.338...14P}. The case for the Plummer halo was challenging. We tested our centroid estimation against the Rockstar halo finder \citep{2013ApJ...762..109B}, and the results of both estimations are compatible. Once the initial and the current snapshot share the same centroid, we proceed to calculate the overdensity maps. The BPM density is a direct code solution. The results are presented in Figure \ref{fig_odensity}, where we show a slice of the overdensity maps in the orbital plane with one cell thickness (1 kpc). From top to bottom: Plummer, NFW1, and NFW2 overdensity maps are presented for (left to right) $t = 0.3,1.0$, and $2.0\,t_\mathrm{dyn}$. At $t = 0.3\,t_\mathrm{dyn}$, the local wake behind the satellite is clearly seen in all models. It is also evident that there is an overdensity and an underdensity dipole (blue and green, respectively). The amplitude of such a dipole is larger for NFW2, followed by NFW1, and finally Plummer. For $t = 1.0\,t_\mathrm{dyn}$, the wake is still visible for all three models, but the relative amplitude regarding the global dipole is getting even for the NFW models. At $t = 2.0\,t_\mathrm{dyn}$, it is hard to see the wake, but the global mode is still clearly detected. Figure \ref{fig_Zoom} shows a zoom in the overdensity map inside the Plummer core radius at $t = 2.0\,t_\mathrm{dyn}$; an over- and underdensity around the satellite is observed, showing that there are modes exchanging angular momentum besides the large scale one. A similar zoom in the NFW models did not show the same structure.

To further study the response of the host halo to the sinking satellite, we calculated the angular power spectrum (APS) of each halo. To that end, we first projected the halo's density map (previously recentered on the initial halo centroid) onto a sphere with a 100 kpc radius. We then calculated the spherical harmonics coefficients \citep[using the python package HEALPY, ][]{2005ApJ...622..759G,2019JOSS....4.1298Z} $c_{\ell,m}$ of the projected density and summed over all $m$ values, for a given $m$: $C_\ell:=\Sigma_m \left|c_{\ell,m}\right|^2$. We refer to $C_\ell$ as a ``mode.'' Figure \ref{fig_APS_Evol} shows the APS, normalized to the initial one, for each host halo and for the three different times. At $t=0.3\,t_\mathrm{dyn}$ (solid lines), the cuspy models are similar, however, the compact NFW2 one (red) seems less responsive, and the modes at $\ell=3$\footnote{\label{APS_DensPart}The analytical method used in our experiments projects the grid cells onto a spherical surface at 100 kpc. The projections of the coordinate planes redistribute the power, shifting the main mode to $\ell=3$; however, if we instead projected only particles, the dominant mode would be $\ell=1,$ in agreement with previous studies such as \citet{2021ApJ...916...55T}.} and $\ell=5$ are similar in amplitude. In contrast, for the NFW1 halo (blue), the $\ell=3$ mode is dominant over the others. The Plummer model (black) seems the most responsive of the three halos based on the APS amplitude. The differences are more dramatic for the following times. At $t=1.0$ (dashed lines) and $2.0\,t_\mathrm{dyn}$ (dotted lines), the APS amplitude of NFW2 grows considerably less than the other models, and the shape becomes smoother, suggesting that the compact and smaller NFW2 halo is less responsive than the NFW1 and the Plummer halos. The latter seems the most responsive because its APS amplitude grows considerably. The APS behavior is in agreement with the overdensity maps. There is a peak at $\ell=3$ and $\ell=5$ for all models, and particularly the Plummer one additionally captures peaks at $\ell=7$, $\ell=9$, and possibly more. At the end of the simulation ($t=2.0\,t_\mathrm{dyn}$), both the NFW1 and the Plummer APS are heavily dominated by the $\ell=3$ mode, while the NFW2 one is less bumpy.

It is also interesting to see the evolution of the bulk velocity map in the orbital plane. Figure \ref{fig_Velocity} shows the maps of the $X$ (upper figure) and $Y$ (lower figure) bulk velocity components for the Plummer, NFW1, and NFW2 halos (top to bottom, respectively), and from $t=0.3,1.0$, and $2.0\,t_\mathrm{dyn}$ (left to right, respectively). The left and central column panels show a negative or positive (blue or red) velocity gradient centered on the satellite illustrating the local wake formation. There is also a global velocity gradient (blue or red). At the rightmost panels, in agreement with the overdensity map, there is no clear signal of the local wake. We may ask if the velocity gradient indicates rotation. In Figure \ref{fig_CStalling.pdf}, we show a zoom into the core region of the bulk velocity components maps (at $t=2.0\,t_\mathrm{dyn}$), now including the $Z$ component. It is evident that the rotation and all angular momentum transfer occur inside the orbital plane. All the density and velocity maps show dependence on the halo velocity profile. Although the results are encouraging, we need to test the robustness of numerical code dependence.

\subsection{Comparison with \emph{N}-body codes}\label{sec_BPMvsNbody}
Besides the convenient properties of BPM handling the shot-noise effects, it is also important to test the reliability of BPM in comparison with standard \emph{N}-body codes. For this reason, we additionally ran some of our numerical experiments with our own PM code and also with a Tree code \citep[Gyrfalcon,][]{2000ApJ...536L..39D}. Figure \ref{BPMTEST} presents the sinking satellite history for the three halo models and the three codes. In general, the three codes capture the same sinking satellite history with differences of around 10$\%$. After 1.5 $t_\mathrm{dyn}$, the sinking stalls for the Plummer model, as well as with all codes, as  reported in previous studies \citep{2015MNRAS.454.3778P}. For the NFW models, the sinking rate also decreases; however, it is close to the resolution.

In addition to testing the consistency of the sinking satellite history across codes, we also tracked the momentum evolution in experiments with the different codes. Figure \ref{Code_Momentum} shows the linear (upper panel) and angular (lower panel) momentum during the $2.0\,t_\mathrm{dyn}$. Indeed, the BPM (solid line) case has lower momentum conservation, likely because it only solves the continuity equation. The rate of change of momentum conservation is significant for BPM but does not grow systematically over time. Hence, the main dynamical effects are consistent with traditional full \emph{N}-body codes, at least for the number of simulated dynamical times, as we can see in Figure \ref{BPMTEST}.

Finally, we may consider whether BPM does indeed hold an advantage in tracing the overdensity signal in the simulations. Figure \ref{Code_SN} shows the overdensity maps (from top to bottom) built based on: BPM, PM, and Tree codes (with two different cloud-in-cell, CIC, resolutions: 1.0 and 0.3 kpc, respectively). For the \emph{N}-body simulations, we used the CIC interpolation to build the density and overdensity maps, while for BPM, we used the numerical solution of the continuity+Poisson equations. Density, in the BPM code, inherits some noise from the velocity estimation; however, it is less noisy than the \emph{N}-body counterparts at a similar grid resolution. The fourth row shows the analysis of the Tree simulation using a CIC grid with equal spatial resolution as the simulation (i.e., 0.3 kpc, Ng=1720). In this case, the satellite or the global mode is hardly seen. However, the structures start to appear when a lower resolution (1.0 kpc) CIC grid is used (third row). The above outlines that the analysis of traditional \emph{N}-body simulations requires special treatment, for example, looking for convergence when smoothing the grid or using adaptive smoothing procedures. The PM case is similar to the Tree smoothed case. Finally, the BPM case directly shows the density and the overdensity fields out of the continuity equation with no smoothing required. All three codes show similar features (see first three rows); however, BPM makes it easy to identify small changes in the density field without any special analysis of the simulations.

Figure \ref{fig_APS} shows the evolution of the APS estimated in simulations of the NFW1 halo using three different codes: BPM, PM, and Tree (solid, dashed, and dotted lines, respectively) with a spatial resolution  of $\sim 1.0$ kpc in the density field. The behavior of the APS is similar between codes: the $\ell=3$ mode has the largest amplitude, while the other odd modes decrease in amplitude as $\ell$ increases. This behavior remains similar for three snapshots. The BPM and PM codes capture the dominant $\ell=3$ mode with the same amplitude, and the Tree code captures the mode with a lower one. For the $\ell=5$ mode, the codes differ in amplitude, from highest to lowest: BPM, PM, and Tree, respectively. The relative amplitude between codes of the $\ell=7$ mode varies over time. For example, at $t=0.3$, BPM has a higher amplitude, but PM and Tree have a lower similar one, while at $t=2.0$, BPM and Tree have equal amplitude, and PM has a lower one.

We may wonder whether the PM code is already capable of capturing the response modes. For that reason, we performed a test with BPM and PM codes using $N_p = 128^3$ particles and a grid of $ N^3_{g} = 512^3$ cells (1 kpc resolution). Figure \ref{fig_NoiseComp} presents the results for the Plummer halo at $t=1.0\,t_\mathrm{dyn}$. The top and middle panels show the overdensity field (built similarly to figure \ref{Code_SN}) for the BPM and PM simulations, respectively. The bottom panel shows the corresponding normalized APS (green lines). We clearly see a significant amount of damping in the PM case (green dashed line); even the relative satellite-halo-center distance is shorter at the scale of $15\%$, suggesting that unresolved modes in the PM code slow down the satellite sinking in the BPM. Increasing the particle number to $256^3$ is not enough (magenta line in figure \ref{fig_NoiseComp}), and possibly $512^3$ may approach BPM; however, the memory cost increment would be a factor of 64 larger for PM in comparison with BPM, considering that the memory cost of both codes is scaled as $N_p + N^3_{g}$.

Regarding the execution time, we made a comparison using a single core. Table \ref{table:comp_cost} summarizes the setup of the simulations and their performance. It is relevant to state that we did not make an exhaustive comparison of code performance. We found that the PM code is $\sim40\%$ faster than the BPM one (for a similar simulation setup). Tree and BPM run at similar times, both codes using the same number of particles. However, BPM easily captures subtle fluctuations in the density field (see Figure \ref{Code_SN}). Furthermore, from Figures \ref{fig_APS} and \ref{fig_NoiseComp}, we can see that the amplitude difference of the mode of order $\ell=3$ between BPM and PM is lower for the simulations with $256^3$ particles (Figure \ref{fig_APS}) than in the case with $128^3$, suggesting that there is a shot-noise effect. If we increase the number of particles to reduce such an effect in the \emph{N}-body simulations, then the memory cost  (and possibly the execution time) would be higher than the BPM simulation (with $256^3$ particles).

\section{Discussion and conclusions} \label{sec:Discusion}
The study of triggered modes in \emph{N}-body simulations of sinking satellites, either to assess their signature in the stellar halo or to study their role in dynamical friction, has been a challenging task requiring large particle numbers and computational resources. Using a new hybrid non-linear CBE+Poisson solver, we successfully detected the overdensity and kinematics response in sinking satellite experiments. We explored the effect on the host halo response related to its density profile. We also discussed the properties of the modes during the core stalling phase.
 
\begin{figure}
    \centering
    \includegraphics[trim=0.3cm 0.1cm 1.2cm 1.1cm, clip, width=0.45\textwidth]{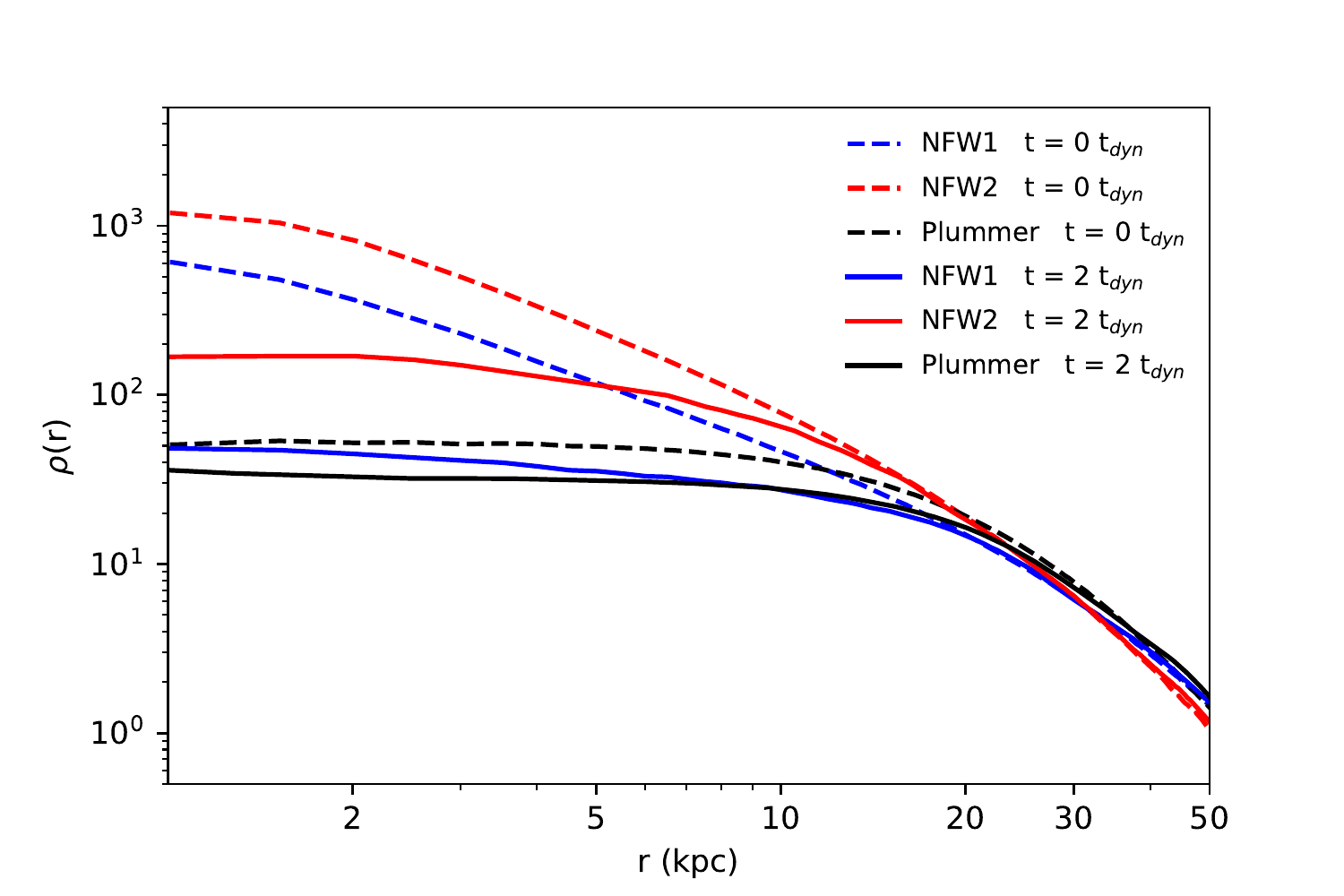}\hfill
    \includegraphics[trim=0.3cm 0.1cm 1.2cm 1.1cm, clip, width=0.45\textwidth]{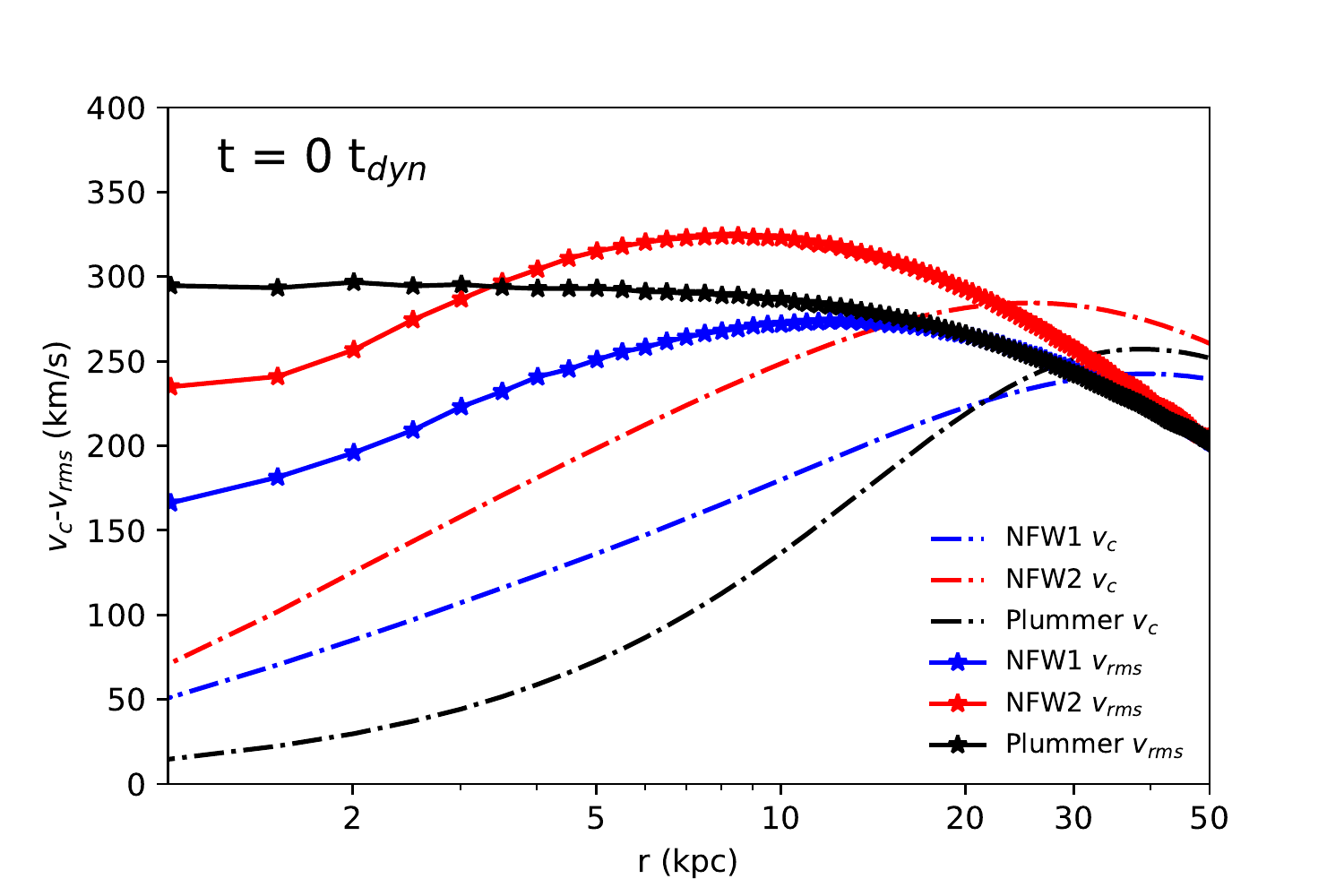} \hfill
    \includegraphics[trim=0.3cm 0.1cm 1.2cm 1.1cm, clip, width=0.45\textwidth]{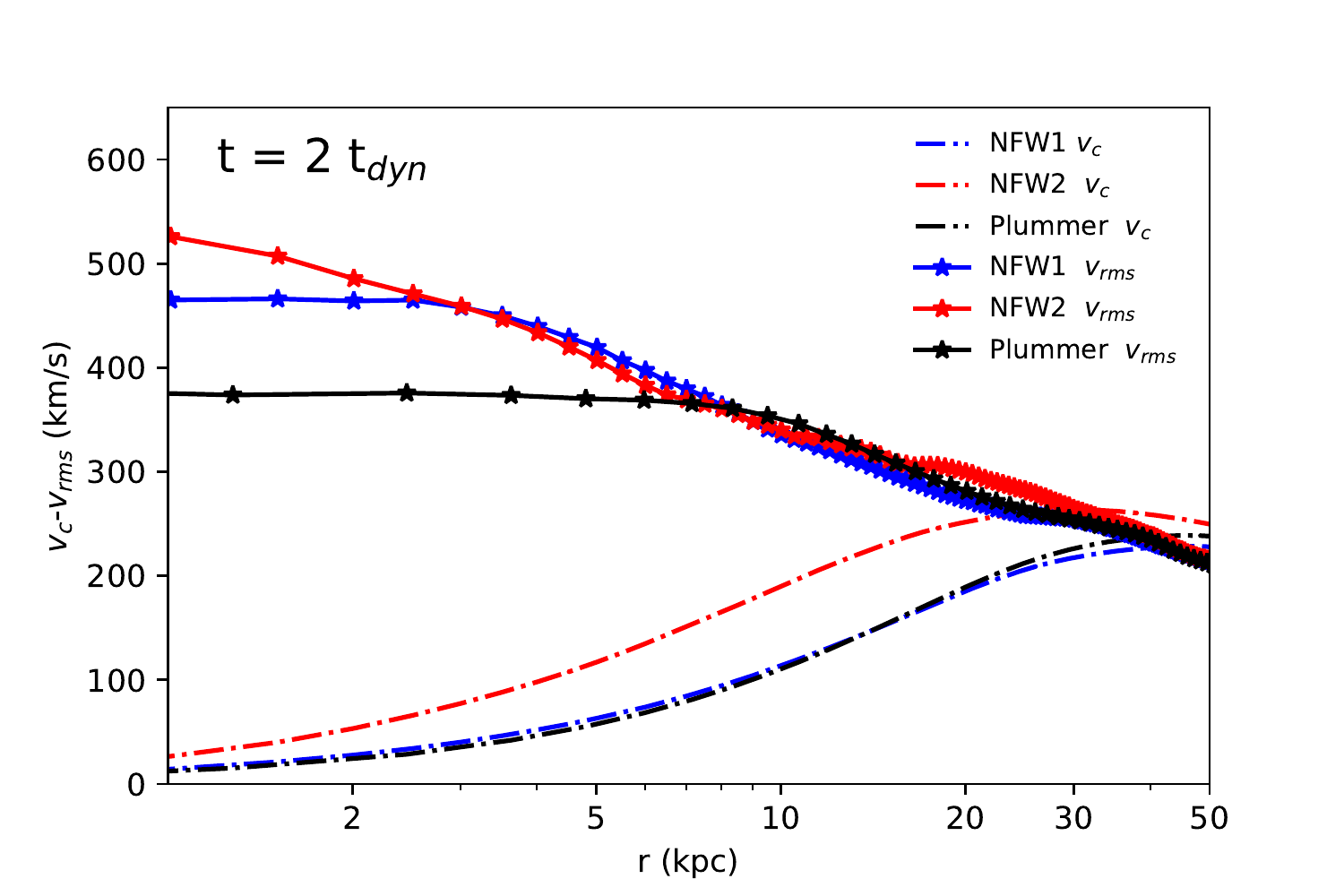}\hfill
    \caption{Core formation. The upper panel shows the initial (dashed) and the final (solid) density profiles for the three halo models. Both NFW models clearly show that a core was developed at 20 (NFW1) and 10 (NFW2) kpc, respectively. The middle panel shows the initial circular velocity (dash-dotted line) and the initial 3D velocity dispersion (solid line with stars). The lowest panel shows the same kinematic profiles at the end of the simulation. \label{fig_CStalling_Vc_Vrms}}
\end{figure}

The upper panel in Figure \ref{fig_CStalling_Vc_Vrms} shows the evolution of the density profile for the host+satellite system assuming the three host halo models (initial correspond to dashed lines and final to solid ones). It clearly shows the formation of a core in the central NFW1 (blue) that turns close to the Plummer (black) density profile. The NFW2 model (red) develops a smaller core. The behavior is not surprising due to the mass ratio between the satellite and the enclosed mass \citep{2010ApJ...725.1707G}. The new code, BPM, presented in this paper correctly captures such a process.

The evolution described above is consistent with the stalled sinking shown in Figure \ref{BPMTEST}. After 1.5 $t_\mathrm{dyn}$, it is possible to observe that the satellite falling holds. Such core stalling phase has been discussed by several studies, such as \citet{2015MNRAS.454.3778P} and the references therein. However, there is no consensus on the explanation behind this effect. In our case, the stalling is observed at around 10 kpc for Plummer and around 2-1 kpc for NFW2 and NFW1.

Figure \ref{fig_Zoom} shows the overdensity modes within the Plummer core. A similar figure in the NFW models shows no structure, possibly because the core radius is smaller and hardly resolved. We may speculate that the central modes in the Plummer core compete with the global mode to weaken dynamical friction. The low amplitude of the density modes in Figure \ref{fig_Zoom} and the high $V_\mathrm{rms}/V_c$ value in Figure \ref{fig_CStalling_Vc_Vrms} \citep[see also][]{2015MNRAS.454.3778P} may contribute to the weakening of dynamical friction \citep{2011MNRAS.416.1181I}, in agreement with recent linear CBE perturbation study \citep[][appeared after our submission]{2021arXiv211210801K}.

\begin{table}
    \caption{Performance of the simulations}
    \label{table:comp_cost}
    \centering
    \begin{tabular}{ c c c c c }
        \hline\hline
        \textbf{Code} & \textbf{$N_p^{1/3}$} & \textbf{$N_g^{1/3}$} & \textbf{SR (kpc)} & \textbf{ET/ET$_\mathrm{BPM}$} \\
        \hline
        BPM & 128 & 512 & 1.0 & 1.0  \\
        PM & 128 & 512 & 1.0 & 0.39 \\
        \hline
        BPM & 256 & 512 & 1.0 & 1.0 \\
        PM & 256 & 512 & 1.0 & 0.40 \\
        Tree & 256 &  & 0.3 & 1.0 \\
        \hline
    \end{tabular}
    \tablefoot{Summary of the simulations' setup and their performance. From left to right, the columns show: code/technique, number of particles, grid size, spatial resolution, and the execution time normalized to the corresponding BPM simulation.}
\end{table}

Our results suggest that the response of the host halo is sensitive to its density profile (see Figure \ref{fig_odensity}). The smaller and compact NFW2 halo shows the strongest amplitude dipole, which is also observed in the flatter APS (Figure \ref{fig_APS_Evol}) compared to the two massive halos, NFW1 and Plummer. All of them show the local wake with a good signal to noise ratio. The evolution of the amplitude and modes of the APS during the simulation (Figure \ref{fig_APS_Evol}) indicates that NFW2 is the least responsive halo, as compared to the Plummer one. This behavior can be understood by looking at the middle panel in Figure \ref{fig_CStalling_Vc_Vrms}, where we show the initial circular velocity (dash-dotted lines) and 3D rms velocity (solid line with stars). It is clear that the NFW2 model (red) has had a higher velocity dispersion since the beginning, possibly blurring the orbital resonances. The lower panel shows the same ($V_c, V_\mathrm{rms}$) curves at the end of the simulation; the behavior of $V_\mathrm{rms}$ is correlated with the core stalling radius. The final stage where core stalling is present is even more different; however, a common feature is that the local wake is not observed.
We summarize our conclusions as follows.

\begin{enumerate}
    \item Our code, named BPM, is able to capture the response modes (local wake, global dipole) in the density and velocity fields with a comparable computational cost to traditional \emph{N}-body simulations. Since BPM is based on resolving a moment hierarchy of the CBE+Poisson equations on a 3D Eulerian grid and using test particles kinematics to close the Boltzmann hierarchy, BPM is less sensitive to shot-noise than the standard \emph{N}-body approach. Therefore, BPM facilitates the detection of subtle fluctuations in the density and velocity fields without special treatment (e.g., higher number of particles, post-processing analysis) that would otherwise be required to achieve the same level of accuracy in a standard \emph{N}-body simulation.

    \item Both the spatial distributions of over- and underdensities and the normalized angular power spectrum depend on the halo density profile, particularly the global dipole strength. The dependence of the overdensity map is more evident when the satellite approaches the core region, and it is related to the central $V_{rms}/V_c$ ratio.

    \item Most of the angular momentum transfer occurs in the orbital plane. At advanced stages of the simulation, we found a central cylindrical rotation (Figure \ref{fig_CStalling.pdf}). In the case of the presence of a stellar component, this may be a sign of past massive accretion events.

    \item During the satellite core stalling stage, there is no clear signature of the local wake; however, low-density contrast modes are detected (Figure \ref{fig_Zoom}). An interesting follow-up project would consist of  studying their contribution to the slower sinking rate.

    \item  Based on the leftmost panels of Figure 3, we conclude that the recent detection of the galactic stellar halo response \citep{2021Natur.592..534C} may contribute to additional constraints on the mass distribution of the dark matter halo in the Milky Way. However, the situation is degenerated both in mass and density profile. 

    \item The capability of BPM for avoiding shot-noise makes it a convenient choice for problems that require resolving low-amplitude modes, with a moderate computational cost in memory. Situations such as the modes discussed in galaxy dynamical problems or low surface brightness structures in galaxies are natural choices. In addition, BPM allows us to study the response modes even in the case of multiple concurrent perturbers.

    \item The right-hand-side term (source) of the Euler-like equation might consider additional dark matter physics such as self-interaction or other effects (equation of state) that produce a possible change in the halo density response (sound speed). An exploration of such an interesting avenue in density and kinematics will be the subject of future studies.
\end{enumerate}

% 
%-------------------------------------------------------------------

\begin{acknowledgements}
      GA acknowledges useful exchange with A. Banerjee. We thank S. Roca-F\'abrega and H. Vel\'azquez for valuable discussions. GA and AT thank support from a CONACyT PhD fellowship. GA, AT, and OV acknowledge support from DGAPA-UNAM grants IN112518,  IG101222,  and AG101620. 
      The authors thank for the facilities of cluster computers:Atocatl/LAMOD-UNAM and Miztli/DGTIC-UNAM. LAMOD is a collaborative project of IA, ICN, and IQ Institutes at UNAM.
\end{acknowledgements}

% 
%-------------------------------------------------------------------

\bibliography{bibfile}

\end{document}